\documentclass[a4paper]{elsarticle}

\makeatletter
\def\ps@pprintTitle{%
 \let\@oddhead\@empty
 \let\@evenhead\@empty
 \def\@oddfoot{}%
 \let\@evenfoot\@oddfoot}
\makeatother

\pdfoutput=1

\usepackage{upgreek}
\providecommand{\e}[1]{\ensuremath{\times 10^{#1}}}
\usepackage{caption} 
\usepackage{amsmath}

\usepackage{fancyhdr}
\fancypagestyle{copyright}{\fancyhf{}\fancyfoot[C]{\copyright 2016. This manuscript version is made available under the CC-BY-NC-ND 4.0 license \url{http://creativecommons.org/licenses/by-nc-nd/4.0/} }\fancyhead[L]{\url{http://dx.doi.org/10.1016/j.apradiso.2016.06.032}}}

\begin{document}

\begin{frontmatter}

\title{A precise method to determine the activity of a weak neutron source using a germanium detector}

\author[secondaddress]{M. J. M. Duke}
\author[mainaddress]{A. L. Hallin}
\author[mainaddress]{C. B. Krauss}
\author[mainaddress]{P. Mekarski\corref{correspondingauthor}}
\ead{mekarski@ualberta.ca}
\author[mainaddress]{L. Sibley}

\address[secondaddress]{SLOWPOKE Nuclear Reactor Facility, University of Alberta, Edmonton, AB T6G 2G7, Canada}
\address[mainaddress]{Department of Physics, University of Alberta, Edmonton, AB T6G 2E1, Canada}

\cortext[correspondingauthor]{Corresponding author}

\begin{abstract}
A standard high purity germanium (HPGe) detector was used to determine the previously unknown neutron activity of a weak americium-beryllium (AmBe) neutron source.  $\gamma$ rays were created through \textsuperscript{27}Al(n,n$'$), \textsuperscript{27}Al(n,$\gamma$) and \textsuperscript{1}H(n,$\gamma$) reactions induced by the neutrons on aluminum and acrylic disks, respectively.  These $\gamma$ rays were measured using the HPGe detector.  Given the unorthodox experimental arrangement, a Monte Carlo simulation was developed to model the efficiency of the detector system to determine the neutron activity from the measured $\gamma$ rays.  The activity of our neutron source was determined to be 307.4~$\pm$~5.0~n/s and is consistent for the different neutron-induced $\gamma$ rays.

\end{abstract}


\end{frontmatter}
\thispagestyle{copyright}

\section{Introduction}

As neutrons are difficult to detect, determining the absolute activity of a neutron source is challenging.  This difficulty increases as the activity of the source decreases.  Sophisticated techniques exist for neutron activity measurements, including the manganese bath technique\cite{MnBath}, proton recoil techniques\cite{ProtonRecoil} and the use of \textsuperscript{3}He proportional counters\cite{3HeCounters}; nevertheless, the development of a method utilizing commonly available high purity germanium (HPGe) detectors would be advantageous.

HPGe's are an industry standard for measuring $\gamma$ ray energies to high precision.  Neutron Activation Analysis\cite{NAA1,NAA2} (NAA) is ordinarily used to determine the elemental composition of materials, as well as to perform neutron flux and activity determinations.  When a target material is exposed to a neutron flux, it results in the activation of previously stable isotopes in the material.  The $\gamma$ rays from the subsequent decay of the unstable activation products may be measured and used to determine the neutron flux or activity.

When determining neutron source activities on the order of 10\textsuperscript{2}~n/s, as in this study, the use of traditional NAA is inadequate\cite{naa_limits}.  The low neutron flux produces insufficient quantities of radionuclides, resulting in levels of radioactivity well below the detection limits of standard $\gamma$ ray spectrometer systems.  This paper presents modifications that enable accurate measurements of the activity of such weak neutron sources by placing the neutron source and targets simultaneously on the HPGe detector combined with use of a Monte Carlo simulation to determine the production rate and detection efficiency of $\gamma$ rays produced from the neutron interactions.

\section{Experimental Setup}

\begin{figure}
\centering
\includegraphics[width=.6\linewidth]{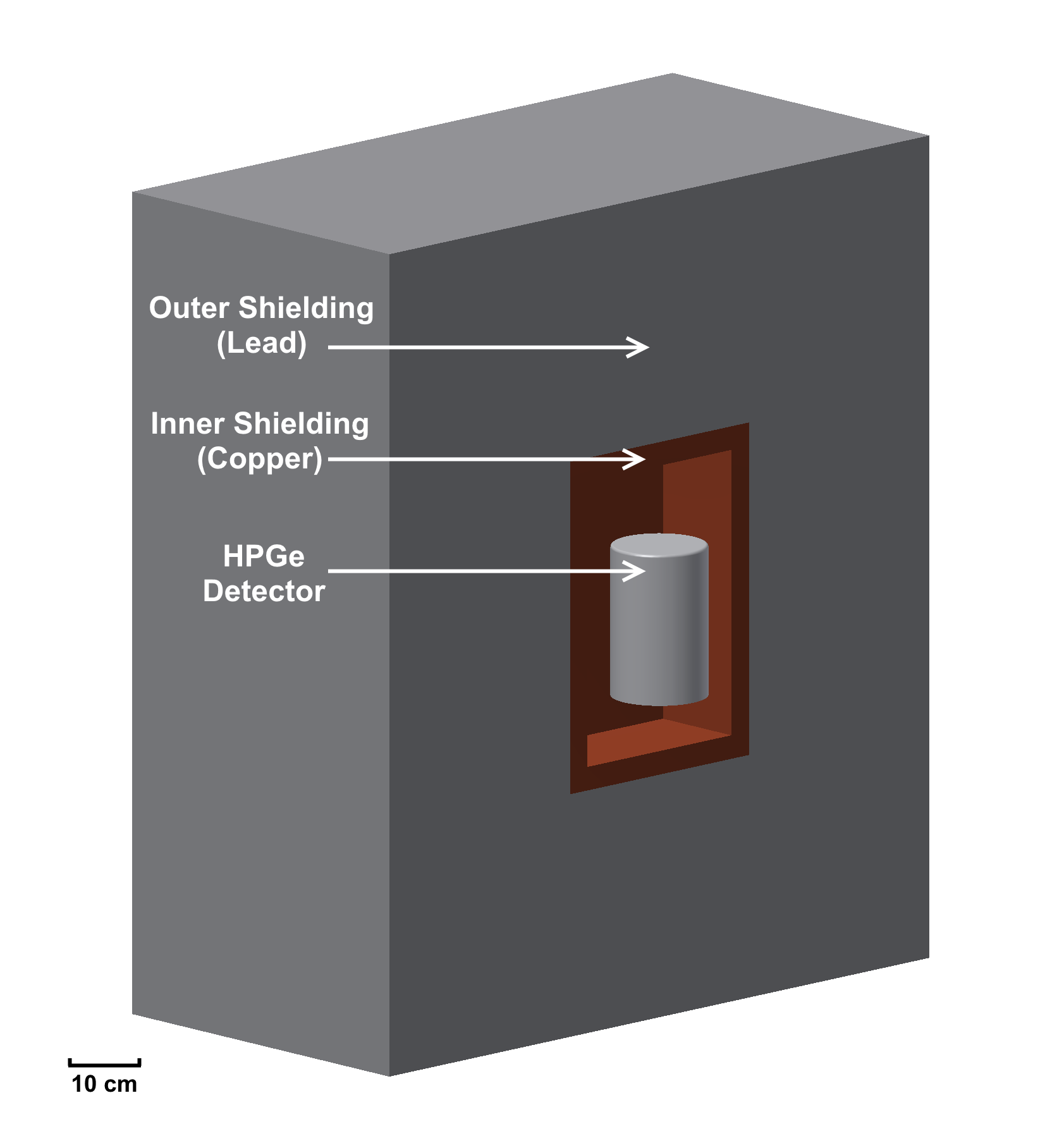}
\caption{Scale representation of the major components of the experimental detector system.  The inner and outer shielding are shown in a cross sectional view.}
\label{fig:Detector}
\end{figure}

The detector system consists of a GC12023 HPGe detector from Canberra Industries.  The crystal is 87.1~mm in diameter by 90.1~mm in length and is located within a 101.6~mm diameter aluminum end cap.  The detector is arranged at the centre of a graded rectangular shield.  The inner shield consists of 25.4~mm thick copper and the outer shield of 250~mm thick lead.  These components are shown in Figure~\ref{fig:Detector}.  A Canberra Lynx digital signal analyzer acquires the spectral data.  The response of the detector is well constrained through calibrations using a standard suite of $\gamma$ ray calibration sources.

The neutron source, made by Eckert \& Ziegler Isotope Products, Inc., is an americium-beryllium (AmBe) AM241SNA02 source.  It consists of a homogeneous mixture of americium oxide and beryllium metal within a double stainless steel encapsulation, each layer with a thickness of 0.8~mm.  The manufacturer-specified activity of the \textsuperscript{241}Am in the AmBe source is 5.5~MBq; however the neutron activity is unknown.  Neutrons with energies ranging up to 11~MeV are produced via the \textsuperscript{9}Be($\alpha$,n) reaction with \textsuperscript{241}Am as the $\alpha$ source \cite{NeutronSpectrum}.

\begin{figure}
\centering
\includegraphics[width=.8\linewidth]{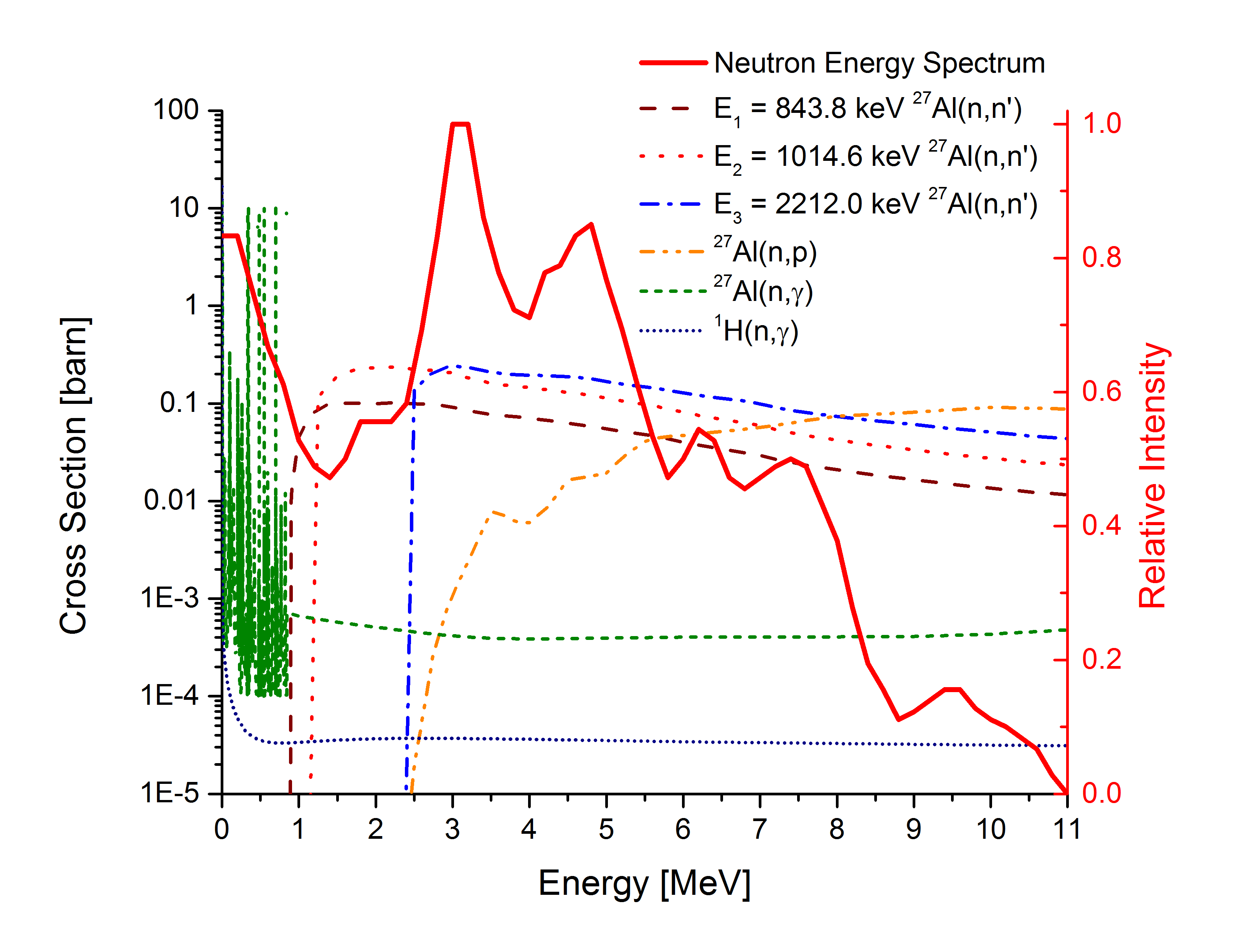}
\caption{Cross sections of the various neutron reactions\cite{ENDF}.  The neutron energy spectrum of an AmBe source is included for reference\cite{NeutronSpectrum}.}
\label{fig:CrossSections}
\end{figure}

Aluminum and acrylic were used as neutron targets.  Aluminum has both fast (\textsuperscript{27}Al(n,n$'$) and \textsuperscript{27}Al(n,p)) and thermal (\textsuperscript{27}Al(n,$\gamma$)) neutron reactions.  Acrylic serves well as a neutron moderator, with its high hydrogen content, and also has a thermal neutron reaction (\textsuperscript{1}H(n,$\gamma$)).  The cross sections for these reactions are shown in Figure~\ref{fig:CrossSections} together with the energy spectrum of neutrons produced from an AmBe source.

The targets were a 17.7~cm diameter by 2.7~cm thick aluminum disk (1.8~kg) and a set of six 19.0~cm diameter by 1.2~cm thick acrylic disks (0.4~kg each), with one disk having a 1.0~cm diameter hole in the centre to accommodate the neutron source, when necessary.  The AmBe source irradiated these targets generating the $\gamma$ rays from the reactions mentioned.  Because of its low activity and the short lifetimes of the reaction products, the neutron source was placed along with the targets in the HPGe detector system.

\subsection{Measurements}

Each measurement (with the AmBe source, target(s), and HPGe) results in a spectrum of gamma ray energies that contain photopeaks at the energies corresponding to specific neutron reactions.  Each peak has three components: 1) $\gamma$ rays originating from the AmBe source neutron interactions in the target materials; 2) background $\gamma$ rays (either inherent in the detector system and AmBe source, or from AmBe source neutron interactions with non-target materials); and 3) background neutrons naturally present in the laboratory interacting with the target materials.  To determine the activity of the AmBe source alone, the contributions from the background $\gamma$ rays and laboratory neutrons must be removed.  This requires 3 measurements per geometric configuration.

In principle, the neutron activity of the AmBe source may be determined from one set of 3 measurements, and by analyzing one specific $\gamma$ ray resulting from one specific neutron reaction.  However, we use a total of 7 geometric configurations and 3 different neutron reactions to more thoroughly evaluate the method presented.  This allows us to later evaluate the consistency in the neutron activities determined by this method and determine a more precise neutron activity through a combined fit.

Table~\ref{tab:geometries} lists the geometric configurations for all measurements performed, each for a period of approximately 24 hours.  Varying the configurations (the geometry) in this way enabled the different thermal and fast neutron reactions to be measured independently.  Each reaction produced distinct peaks in the $\gamma$ ray energy spectrum.  The count rate of any particular peak, $A_{reaction}$, is determined by:
\begin{equation}
A_{reaction} = A_{meas} - A_{bkgd} - A_{lab} .
\label{eq:reaction_activity}
\end{equation}

\begin{figure}
\centering
\includegraphics[width=1.\linewidth]{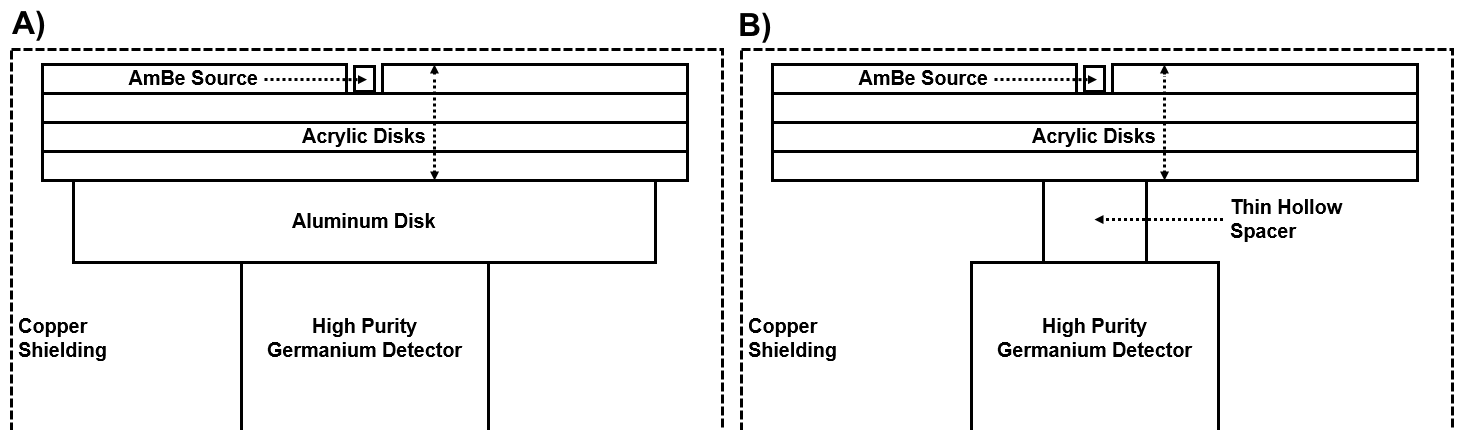}
\caption{Experimental setup of geometric configuration 5.  A) shows the activation and measurement of the aluminum disk.  B) shows the corresponding background measurement.  The number and type of disks were varied.  These target materials were always placed between the source and detector.}
\label{fig:ExperimentalSetup}
\end{figure}

\noindent $A_{meas}$ is the peak count rate in a configuration with the target and source, $A_{bkgd}$ with the source only, and $A_{lab}$ with the target only.  These correspond to columns 2, 3 and 4 of Table~\ref{tab:geometries}, respectively, and each represents a separate measurement taken on the HPGe detector.  An example diagram of the experimental setup used to measure $A_{meas}$ is shown in Figure~\ref{fig:ExperimentalSetup}A.

$A_{bkgd}$ consists of the $\gamma$ ray backgrounds (dominated by those originating from the AmBe source) including those resulting from activation of non-target components of the detector system, such as the aluminum end cap of the HPGe.  These measurements were done in a way such that the only volume missing was the target volume.  In the case of the aluminum disk geometries (configurations 1--5 in Table~\ref{tab:geometries}), the disk was removed and replaced with thin spacers to elevate the remaining acrylic disks to maintain the neutron source geometry, an example of which is shown in Figure~\ref{fig:ExperimentalSetup}B.  For the acrylic disk geometries (configurations 6 and 7 in Table~\ref{tab:geometries}), the disks were removed and the neutron source was placed on a spacer only.

$A_{lab}$ is the contribution from background neutrons in the laboratory, such as those from cosmogenic sources.  These rates were determined by performing the measurements without the neutron source present.  A small contribution was observed from the \textsuperscript{1}H(n,$\gamma$) reaction in geometries 6 and 7 (corresponding to 1.2\%--1.6\% of $A_{meas}$).  The effect of $A_{lab}$ was negligible for all of the other reactions across the series of geometries.

\begin{table*}[t]
  \tiny
  \centering
    \caption{Geometric configurations of the measurements performed.  The disks and source were placed on top of the HPGe detector in the order listed.  Three measurements are made for each geometry ID in order to isolate the contribution from reactions on the target alone.}
    \begin{tabular}{c|c|c|c}
    \textbf{ID} & \textbf{Measurement} & \textbf{Background from AmBe} & \textbf{Background from laboratory} \\
    \hline
    1     & Al disk, AmBe & Spacer, AmBe & Al disk \\
    2     & Al disk, 1 acrylic disk, AmBe & Spacer, 1 acrylic disk, AmBe & Al disk, 1 acrylic disk \\
    3     & Al disk, 2 acrylic disks, AmBe & Spacer, 2 acrylic disks, AmBe & Al disk, 2 acrylic disks \\
    4     & Al disk, 3 acrylic disks, AmBe & Spacer, 3 acrylic disks, AmBe & Al disk, 3 acrylic disks \\
    5     & Al disk, 4 acrylic disks, AmBe & Spacer, 4 acrylic disks, AmBe & Al disk, 4 acrylic disks \\
    6     & 5 acrylic disks, AmBe & Spacer, AmBe & 5 acrylic disks \\
    7     & 6 acrylic disks, AmBe & Spacer, AmBe & 6 acrylic disks \\
    \end{tabular}%

  \label{tab:geometries}%
\end{table*}%

\subsection{Damage Considerations}

This prompt gamma NAA (PGNAA) approach, with its simultaneous activation and measurement, exposes the HPGe crystal to a small neutron flux which may potentially cause damage and result in resolution degradation.  The total neutron exposure of the HPGe crystal during this experiment can be estimated by multiplying the neutron flux through the solid angle subtended by the detector by the exposure time in that configuration.  Summing over all of the measurements performed, this exposure was estimated to be at or below 10\textsuperscript{6}~n/cm\textsuperscript{2}.  Noticeable damage is seen to occur after total neutron exposures on the order of 10\textsuperscript{9}--10\textsuperscript{10}~n/cm\textsuperscript{2}~\cite{Damage}.  No noticeable deterioration of the HPGe detector's performance was observed.

\section{Simulation}

In order to determine the number of neutrons required to produce a detection in the peak region of the $\gamma$ ray energy spectrum, a Monte Carlo simulation was used.  In the case of simple geometric setups and $\gamma$ ray interactions alone, standard calibration sources are used to determine this key factor.  However, this is not possible in the case of complex geometries and with the inclusion of neutron interactions.

Simulations were made using Geant4\cite{Geant4}.  Geant4 tracks particles traveling through matter, models their interactions and allows the user to extract information of interest.  To simulate an experiment, the user must define the geometry, physics processes and data handling during tracking.  Critical components of the Geant4 simulation are described below.  These simulations were used to convert the measured reaction rates to neutron source strengths.

\subsection{Geometry}

Each experimental geometric configuration was modeled in detail, including the copper and lead shielding, HPGe, aluminum and acrylic plates, and neutron source.  The aluminum target was defined with the elemental composition of the aluminum alloy used (6061), which includes small amounts of magnesium, silicon, iron, copper, chromium, and trace amounts of other elements.  The aluminum concentration in the alloy was determined by conventional Instrumental Neutron Activation Analysis (INAA) at the University of Alberta SLOWPOKE Nuclear Reactor Facility and found to be 97.4~$\pm$~0.4~wt.~\%.

Neutrons were generated isotropically from positions distributed uniformly within the central powder of the AmBe geometry.  The energies of the neutrons were drawn from the distribution in Figure \ref{fig:CrossSections}.

\subsection{Physics}

Within Geant version 4.10.0, the ``Standard Electromagnetic" package was used to handle the $e^{-}$ and $\gamma$ ray transport and interactions.  The ``Radioactive Decay" physics list was used to model the decay of the isotopes generated through neutron interactions.  This includes the generation of all decay products ($e^{-}$'s, $\gamma$ rays, metastable isotopes, etc.) up until a stable isotope state is reached.  The ``QGSP\_BIC\_HP" hadronic physics list uses the ``High Precision" neutron cross sections to handle the neutron transport and reactions in the nominal 0--10~MeV energy range.  The inelastic scattering process was overridden to produce excited aluminum atoms which were then handled by the radioactive decay processes to produce $\gamma$ rays.  This procedure resulted in $\gamma$ ray spectra consistent in peak width and location with measured spectra.

\subsection{Data Handling}

An event is defined as the generation of one neutron along with all of the secondary particles it produces along its trajectory through the simulated experimental geometry.  Several different values are extracted and recorded during each of the events.  Most importantly, the total energy deposited in the HPGe detector (primarily from $\gamma$ rays) for each neutron generated is recorded.

10\textsuperscript{7} neutrons ($N_{simulated}$) were simulated for each experimental geometry.  A $\gamma$ ray spectrum was created by summing over the total energy deposited in each event.  The same analysis algorithm was used to determine the integrals under the peaks of the simulated spectra as was used for the spectra produced from measurements.  Integrating under the peaks ($N_{peak}$) in these simulated spectra, the neutron source strength ($A_{n}$) was determined from the measured data as:

\begin{equation}
A_{n} = A_{reaction} \frac{N_{simulated}}{N_{peak}}.
\label{eq:neutron_activity}
\end{equation}

\noindent The ratio $N_{peak}/N_{simulated}$ represents the absolute system efficiency, which ranged from 1.57\e{-5}--4.94\e{-4}, depending on the $\gamma$ ray energy.

\section{Calibration}

How accurately the simulation reproduces measured $\gamma$ ray spectra was evaluated by comparing the simulated efficiency of detecting $\gamma$ rays produced by a radioactive source to the known efficiency of the HPGe detector for detecting $\gamma$ rays from a calibration source containing the same radionuclide.  The most precise calibration source used for this comparison was a \textsuperscript{60}Co point source with an activity known to 1.9~\% uncertainty.  \textsuperscript{60}Co emits $\gamma$ rays at 1173~keV and 1332~keV (which may sum to 2505~keV) which are in the energy region of the $\gamma$ rays produced by the target neutron reactions.

\begin{figure}[h]
\centering
\includegraphics[width=.8\linewidth]{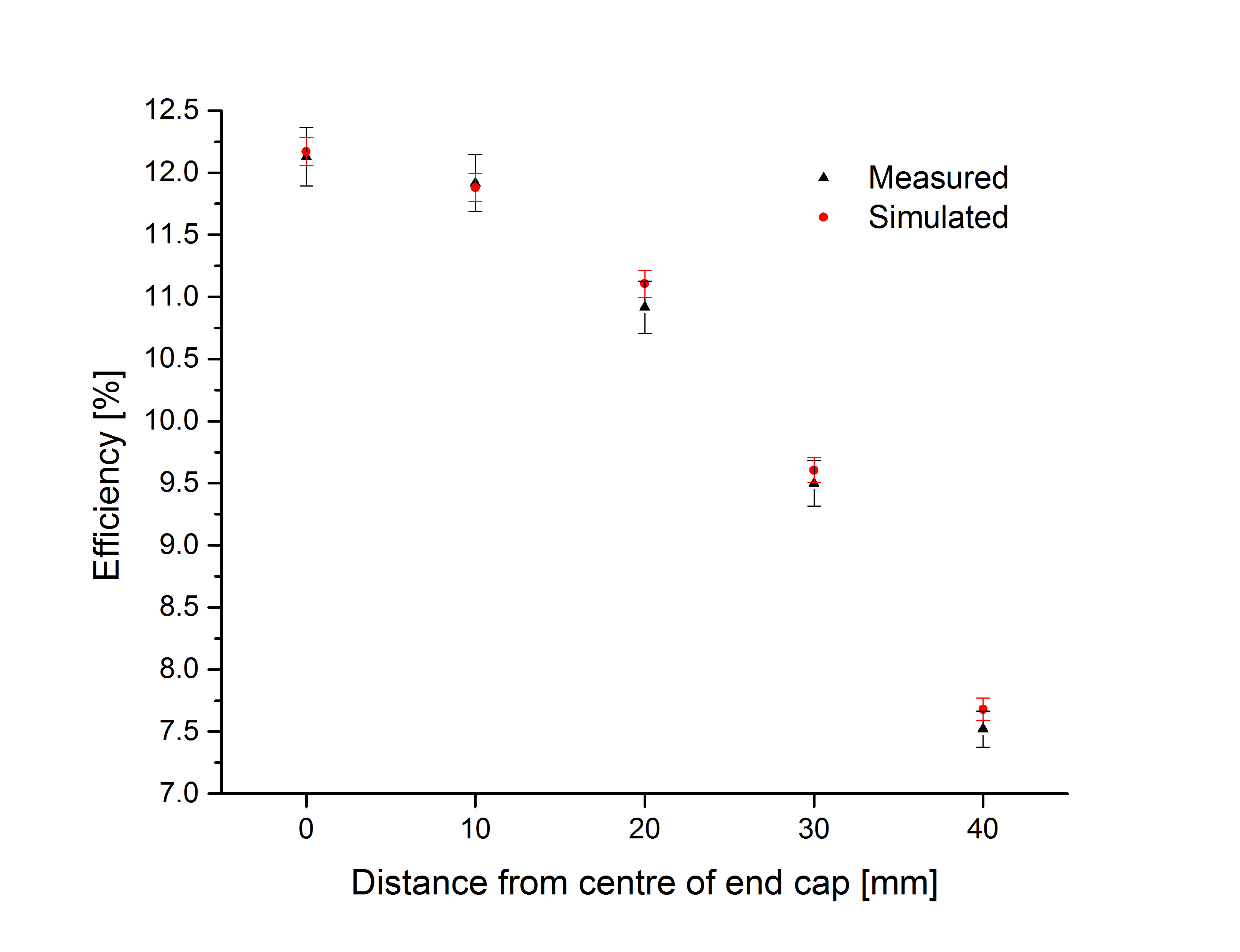}
\caption{Measured and simulated efficiencies for detecting \textsuperscript{60}Co through either of the 1173~keV photopeak, 1332~keV photopeak, or 2505~keV sum peak.  The efficiency is shown for varying distances of the calibration source from the centre of the detector's end cap.}
\label{fig:efficiencies}
\end{figure}

This \textsuperscript{60}Co source was also moved radially from the centre of the end cap to investigate the change in efficiency over the crystal surface (and to gauge the accuracy of the modelled internal geometry of the HPGe detector).  In all cases, the efficiency determined from the simulation was within the uncertainty of the measured efficiency from the HPGe detector (dominated by the uncertainty of the calibration source activity).  This efficiency comparison is shown in Figure~\ref{fig:efficiencies}.  As a result, we conclude that the simulation models the total efficiency of measuring $\gamma$ rays with this specific detector to a precision of 1.9~\%.  This is taken into account in later neutron activity determinations.

\section{Results and Discussion}

Performing the series of measurements described in Table~\ref{tab:geometries}, peaks corresponding to the $\gamma$ rays from the target reactions were observed.  These will be discussed in detail and presented alongside the associated simulated measurements.

\subsection{Fast Neutron Reactions}
\label{sec:correction}

\begin{figure}[h]
\centering
\includegraphics[width=.8\linewidth]{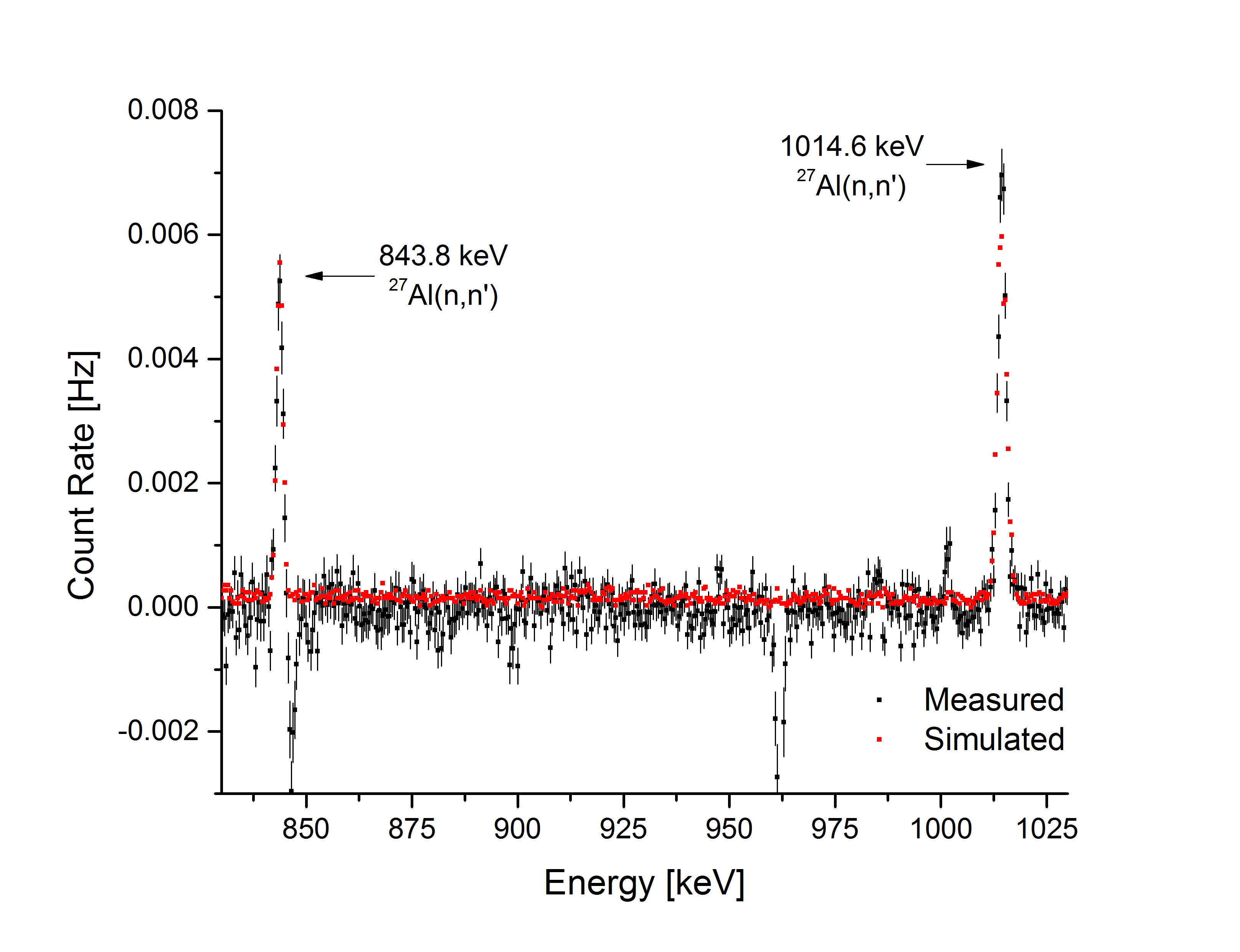}
\caption{Measured and simulated energy spectra for the counting in geometric configuration 5 (four acrylic disks on top of the large aluminum disk). The peaks seen are the result of the \textsuperscript{27}Al(n,n$'$) reaction.}
\label{fig:843_1014Spectrum}
\end{figure}

The \textsuperscript{27}Al(n,n$'$) reaction, which dominates in the fast neutron regime, was observed when the aluminum disk was present.  The 844~keV and 1015~keV peaks, corresponding to the 1\textsuperscript{st} and 2\textsuperscript{nd} energy level transitions of \textsuperscript{27}Al, are shown in Figure \ref{fig:843_1014Spectrum} and are compared to the simulation.  The peak corresponding to the 3\textsuperscript{rd} energy level transition at 2212~keV was also observed in the data at a much lower activity.  The negative peaks observed in the measured spectrum result from the background subtraction, where the removal of material results in less attenuation of background peaks.

The \textsuperscript{27}Al(n,p) reaction contributes to the 844~keV and 1015~keV peaks.  Simulations showed that this contribution is small (2.0\%--3.0\% for the different geometries).  This contribution is taken into account in the subsequent analysis.

From evaluation of the time between interactions in the simulation, the fast neutron reactions occur $<$10~ns after neutron emission.  For AmBe sources, a 4438~keV $\gamma$ ray is sometimes emitted along with the neutron.  This occurs for (57.5~$\pm$~2.8)~\% of the neutrons produced\cite{GammaRatio}.  This $\gamma$ ray may also deposit energy in the HPGe detector in coincidence with an aluminum $\gamma$ ray.  Simulating this 4438~keV $\gamma$ ray provided a coincidence correction to the total efficiency applied to a given photopeak.  This correction for the \textsuperscript{27}Al(n,n$'$) was small, ranging from 3.2--6.1~\%.

\subsection{Thermal Neutron Reactions}

\begin{figure}
\centering
\includegraphics[width=.8\linewidth]{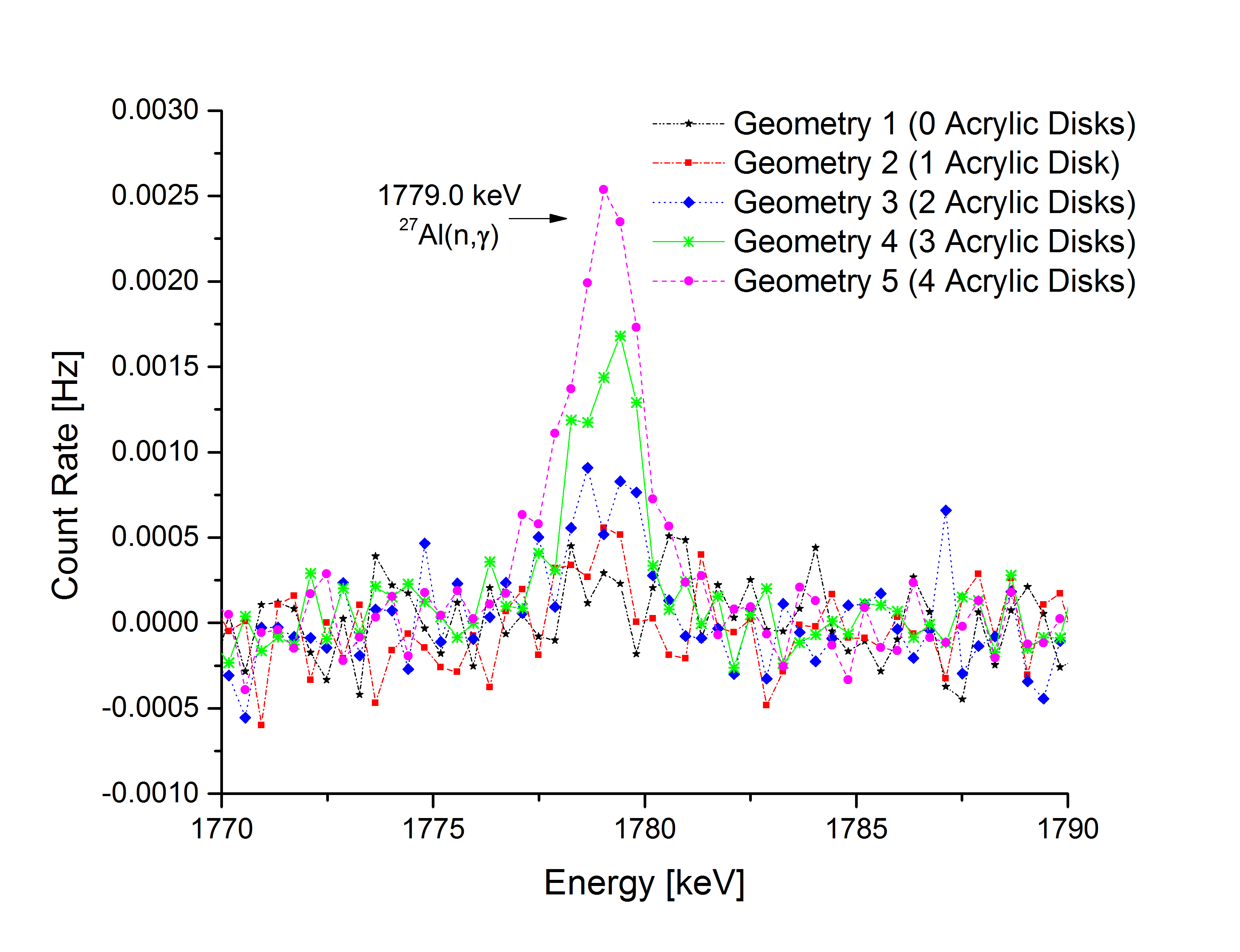}
\caption{Measured \textsuperscript{27}Al(n,$\gamma$) peak for different counting geometries.  An increasing peak activity is seen for increased neutron moderation by the acrylic (error bars removed for clarity).}
\label{fig:1778SpectrumCompare}
\end{figure}

Peaks associated with the \textsuperscript{27}Al(n,$\gamma$) and \textsuperscript{1}H(n,$\gamma$) reactions were observed for experimental geometries 3--7.  Figure \ref{fig:1778SpectrumCompare} illustrates these results and also demonstrates the effectiveness of acrylic as a neutron moderator.

\begin{figure}
\centering
\includegraphics[width=.8\linewidth]{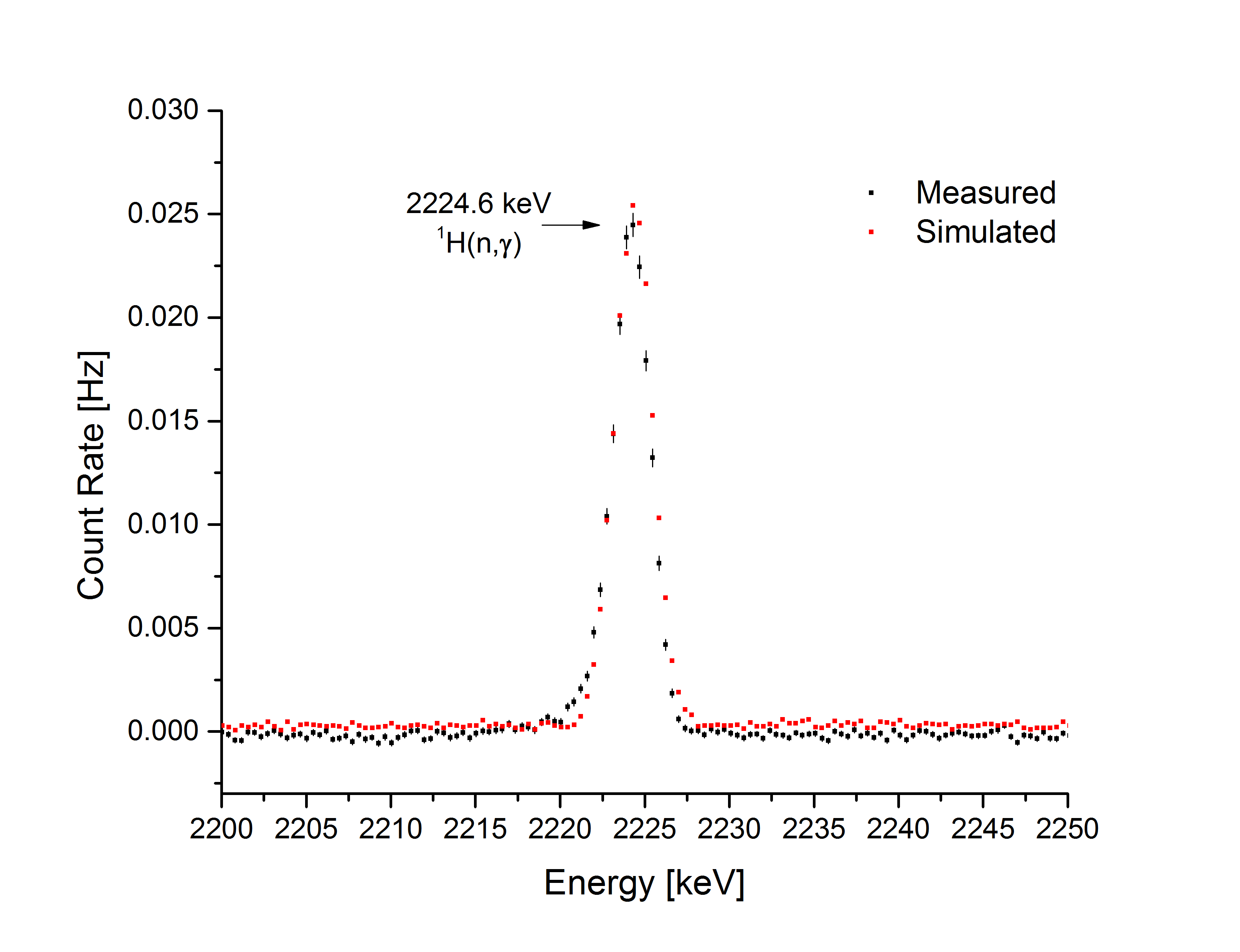}
\caption{Measured and simulated energy spectra for geometric configuration 7 (six acrylic disks). The peak observed is the result of the \textsuperscript{1}H(n,$\gamma$) reaction.}
\label{fig:2224Spectrum}
\end{figure}

The \textsuperscript{1}H(n,$\gamma$) 2225~keV peak was seen in all spectra from all geometries with acrylic.  A peak from geometry 7 is shown in Figure~\ref{fig:2224Spectrum}.  The 4438~keV $\gamma$ ray correction was negligible for thermal neutron reactions.   This is due to the fact that the approximate time between the \textsuperscript{1}H(n,$\gamma$) and \textsuperscript{27}Al(n,$\gamma$) reactions and neutron emission was {\raise.17ex\hbox{$\scriptstyle\mathtt{\sim}$}}10--50~$\upmu$s (determined through interaction times in the simulation).  The $\gamma$ rays from these reactions are then detected separately in the HPGe detector from the 4438~keV $\gamma$ ray emitted from the AmBe source.

\subsection{Activity Determination}
\label{sec:activity}
\begin{figure}
\centering
\includegraphics[width=.8\linewidth]{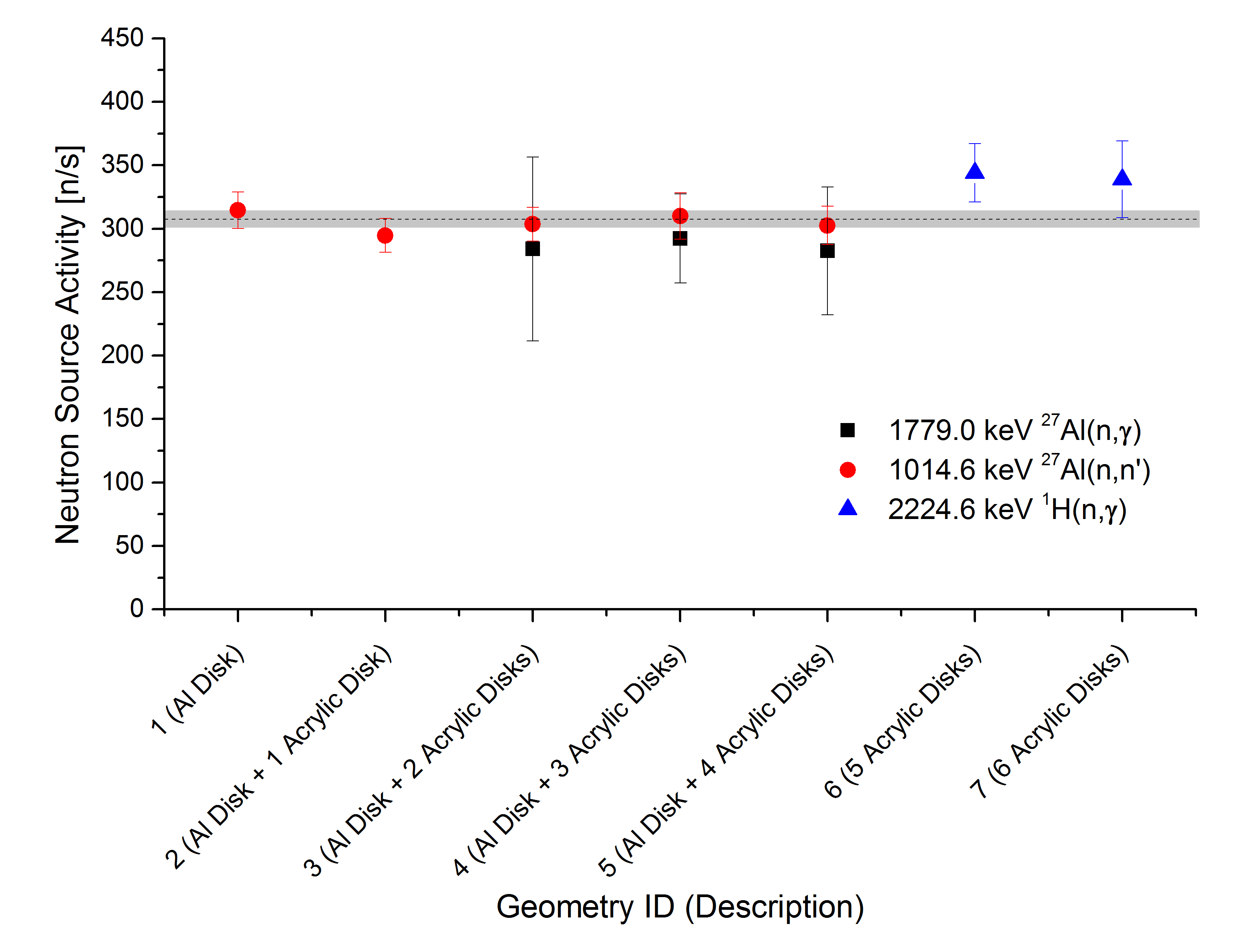}
\caption{Neutron source strengths extracted from the three reactions.  The dashed line is fit to the data and corresponds to a neutron activity of 307.4~$\pm$~5.0~n/s.}
\label{fig:ActivityPlot}
\end{figure}

By comparing a measurement to its associated simulation, a neutron source activity was calculated for each reaction observed in each geometry.  For the \textsuperscript{27}Al(n,n$'$) reaction, the interference-free 1015~keV peak was used for activity calculations though multiple peaks were observed.  From the full system efficiencies determined by the simulations, the neutron source activities were calculated from Equations~\ref{eq:reaction_activity} and \ref{eq:neutron_activity}.  The results are shown in Figure \ref{fig:ActivityPlot}.  A best fit to these data yields a neutron activity of 307.4~$\pm$~5.0~n/s.

\subsection{Systematics}

A further series of measurements were performed to quantify the level of systematic uncertainty present in this experiment.  The contributions of the different major systematics evaluated are quantified in Table~\ref{tab:systematics}.  One dominant systematic uncertainty resulted from the placement of the different targets and neutron source with respect to the HPGe (an uncertainty of 5~mm from nominal centre was taken in radial position).

\begin{table*}[t]
  \small
  \centering
    \caption{Effect on the various systematic evaluated.  The range corresponds to these effects over the applicable geometric configurations which result in the reactions listed.}
    \begin{tabular}{c|c}
    \textbf{Systematic} & \textbf{Effect on neutron activity}\\
    \hline
    Source/target positioning \textsuperscript{27}Al(n,$\gamma$) & 5.3--19.1\% \\
    Source/target positioning \textsuperscript{1}H(n,$\gamma$) & 5.9--6.2\% \\ 
    Source/target positioning \textsuperscript{27}Al(n,n$'$) & 1.8--3.3\% \\
    Neutron spectrum \textsuperscript{27}Al(n,$\gamma$) & 0.8--3.0\% \\
    Neutron spectrum \textsuperscript{1}H(n,$\gamma$) & 0.8--2.9\% \\
    Neutron spectrum \textsuperscript{27}Al(n,n$'$) & 0.1--3.2\% \\
    \end{tabular}%
  \label{tab:systematics}%
\end{table*}%

Further evaluation quantified the uncertainty resulting from our knowledge of the neutron source spectrum.  Small shifts (up to 10\%) were added to different parts of the source spectrum with simulations being repeated.  The subsequent effect was seen to be non-negligible but smaller than that of positioning.  The total uncertainties attributed to these systematic effects are included along with the statistical uncertainties in Figure~\ref{fig:ActivityPlot}.

\subsection{Sensitivity}

For each peak measured in all geometries, the minimum detectable activity (MDA) was determined from the corresponding background measurements.  From the aluminum reactions, geometry 1 was observed to have the lowest MDA (specifically from the \textsuperscript{27}Al(n,n$'$) 1015~keV reaction).  This activity corresponds to a neutron source activity of 9.5~$\pm$~0.4~n/s.  Similarly for acrylic, geometry 7 had the lowest MDA, which corresponds to a neutron activity of 4.2~$\pm$~0.3~n/s.  These sensitivities are well below the measured neutron activity of the AmBe source.

\subsection{Activity Determination from the 4438 keV $\gamma$ Ray}

\begin{figure}
\centering
\includegraphics[width=.8\linewidth]{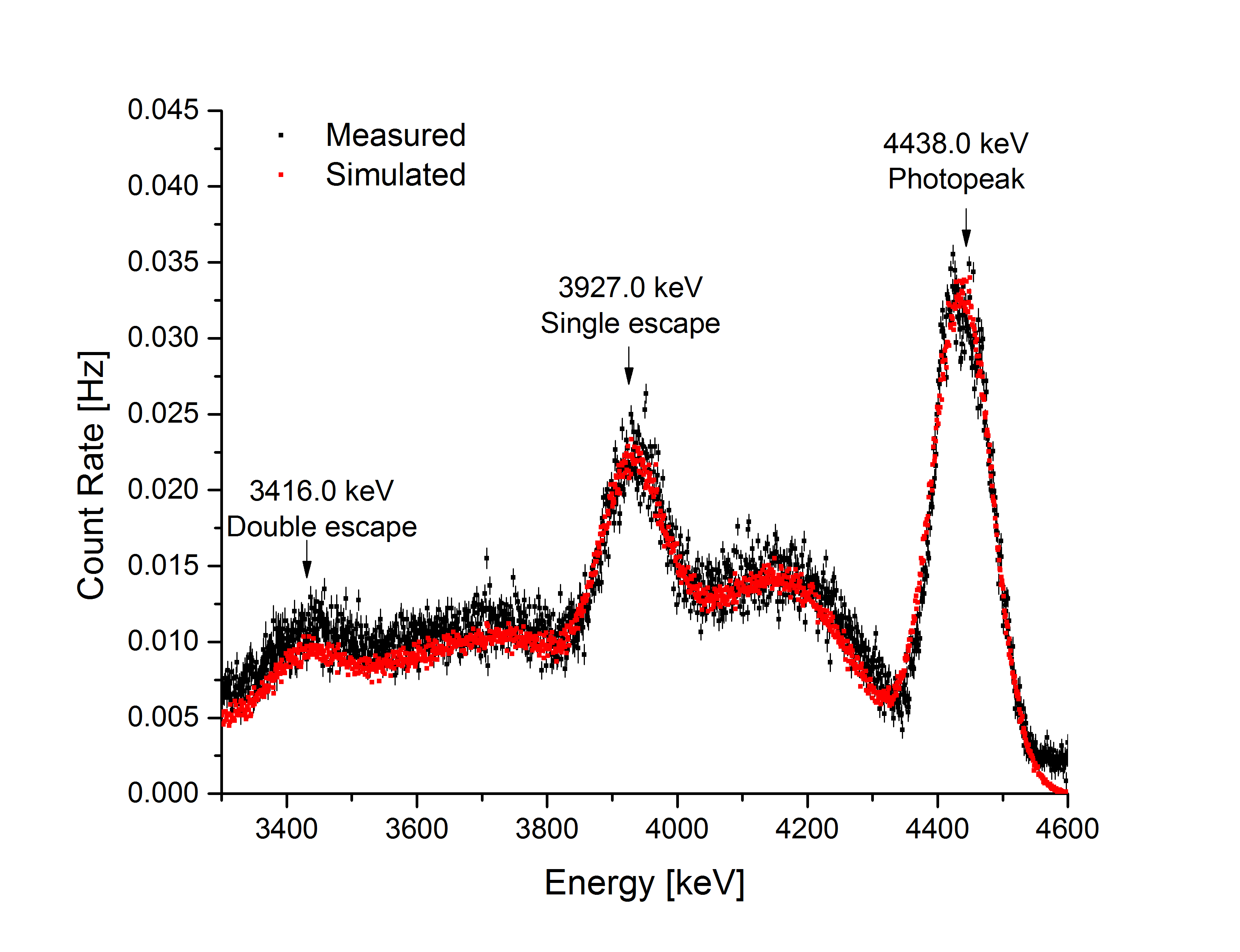}
\caption{Measured and simulated energy spectra for the AmBe source when placed directly on the detector.  The 4438~keV peak is seen along with the associated single and double escape peaks.}
\label{fig:4438Spectrum}
\end{figure}

The neutron activity was also determined in a separate analysis by directly measuring the 4438~keV $\gamma$ ray emitted by the neutron source.  To obtain a photopeak detection efficiency, the same simulation program was used to model the detection of these $\gamma$ rays.  Figure~\ref{fig:4438Spectrum} compares the measured and simulated spectra.  From the peak activity, the neutron activity of the AmBe source was determined to be 305.0~$\pm$~16.1~n/s, consistent with the results of Section~\ref{sec:activity}.  This relies on knowing the $\gamma$ ray to neutron ratio, which we take again to be 0.575~$\pm$~0.028~\cite{GammaRatio}.  

As mentioned in Section~\ref{sec:correction}, using the $\gamma$ rays from the \textsuperscript{27}Al(n,n$'$) reaction (1014.6~keV) to determine the neutron activity also included the use of this ratio.  In determining the neutron activity from these $\gamma$ rays, the ratio is used as a small coincidence correction on the measurement efficiencies. While this makes these two analyses not independent, the effect of the correction based on the $\gamma$ ray to neutron ratio is small (3.2--6.1~\%).  On the contrary, determining the neutron activity from the 4438~keV $\gamma$ rays relies entirely on the accuracy of this value.  The other reactions used to determine the neutron activity are independent of this ratio.  

The large uncertainty in the gamma to neutron ratio leads to a less precise measurement of the neutron source activity than that from the analysis of the neutron reactions on the targets. As a result, we quote the latter as our measured value and use the former only as a tool for comparison.

\section{Conclusions}

A method was created to determine the previously unknown strength of a weak neutron source using a HPGe detector in conjunction with a Monte Carlo simulation.  Three different reactions in both the thermal and fast neutron regimes were used to determine the neutron activity of an AmBe source.  The independently-determined activities from a series of geometric configurations show consistency and result in a neutron activity of 307.4~$\pm$~5.0~n/s.

\section*{Acknowledgments}

This work was supported by the Natural Sciences and Engineering Research Council of Canada, Canada Foundation for Innovation, Alberta Innovates -- Technology Futures, Alberta Innovation \& Advanced Education, and the Killam Trust.

\bibliography{neutronBib}

\end{document}